\documentclass[prl,twocolumn]{revtex4}
\usepackage{epsfig}
\usepackage{bm}
\usepackage{amsmath}

\begin{document}

\title{Role of correlation integrals in helical isotropic turbulence}

\author{A. Bershadskii}

\affiliation{
ICAR, P.O. Box 31155, Jerusalem 91000, Israel
}

\begin{abstract}
 It is shown that the correlation integrals (invariants) play an important, and at certain conditions a dominant, role in helical isotropic homogeneous turbulence both in the inertial and near-dissipation ranges. Results of direct numerical simulations and recent laboratory experiments with multiscale grids have been used for this purpose. A possibility of spontaneous breaking of reflection symmetry has been also briefly discussed in this context. 
  
\end{abstract}

\maketitle

\section{Correlation integrals}

  Correlation integrals were introduced into hydrodynamics by L.G. Loitsianskii (see, for instance, Ref. \cite{my} and references therein). Namely, the Loitsianskii velocity-velocity correlation integral for isotropic and homogeneous motion is
$$
\mathcal{L} = - \int r^{2} \langle {\bf u} ({\bf x},t) \cdot  {\bf u} ({\bf x} + {\bf r},t) \rangle d{\bf r}  \eqno{(1)}
$$
 For certain class of the initial conditions this integral is a finite invariant of the Navier-Stokes (viscous) equations for incompressible fluids. Then G. Birkhoff introduced an additional correlation integral (see, for instance, Ref. \cite{saff} and references therein). The Birkhoff-Saffman integral for isotropic and homogeneous motion is 
$$
\mathcal{S} = \int  \langle {\bf u} ({\bf x},t) \cdot  {\bf u} ({\bf x} + {\bf r},t) \rangle d{\bf r}  \eqno{(2)}
$$
For certain class of the initial conditions this integral is also a finite invariant of the Navier-Stokes (viscous) equations for incompressible fluids.\\

   While the invariance of the Loitsianskii integral is related to the angular momentum conservation (or to rotational symmetry - a consequence of the Noether's theorem), the invariance of the Birkhoff-Saffman integral is related to the linear momentum conservation (or to the spatial translational symmetry - also a consequence of the Noether's theorem). The spatial rotational and translational symmetries correspond to spatial isotropy and homogeneity, respectively (in this paper we understand the term 'isotropy' in its narrow meaning - rotational symmetry only). Therefore these invariants have a fundamental nature. \\
   
     The velocity correlation integrals were originally introduced in order to study behaviour of kinetic energy spectrum at small values of wavenumber $k$
$$
E(k) \propto \mathcal{S}k^2  \eqno{(3)}   
$$     
or 
$$
E(k) \propto \mathcal{L}k^4  \eqno{(4)}   
$$     
when the Birkhoff-Saffman integral $\mathcal{S}$ can be considered as a negligible one.  \\

  It was recently shown \cite{otto1} that there is an additional, now velocity-vorticity, correlation integral   
$$
\mathcal{C} =\int \langle {\bf u} ({\bf x},t) \cdot  {\boldsymbol \omega} ({\bf x} + {\bf r}, t) \rangle d{\bf r}   \eqno{(5)}
$$
that is an invariant of the Navier-Stokes (viscous) equations for the incompressible fluids in the isotropic homogeneous case (${\boldsymbol \omega} ({\bf x}, t) = \nabla \times {\bf u} ({\bf x},t)$ is the vorticity field). \\

 Generalization of the large-scale (small $k$) kinetic energy spectrum Eqs. (3)-(4) for the Chkhetiani invariant $\mathcal{C}$ dominated motion is 
$$
E(k) \propto |\mathcal{C}|k \eqno{(6)}   
$$     

   Obviously the Chkhetiani invariant $\mathcal{C}$ can be related to the well known helicity \cite{otto1}
$$
\mathcal{H} = \int  h({\bf x},t)~ d{\bf x}  
$$
where the helicity density 
$$
h ({\bf x},t) =   {\bf u} ({\bf x},t) \cdot  {\boldsymbol \omega} ({\bf x}, t)   \eqno{(7)}
$$

  Another correlation integral related to the helicity is the Levich-Tsinober integral
\cite{lt}-\cite{l}
$$   
I = \int  \langle  h ({\bf x},t) \cdot   h ({\bf x} + {\bf r},t) \rangle d{\bf r}  = \lim_{V\to\infty} \frac{1}{V} \langle \mathcal{H}^2 \rangle \eqno{(8)}
$$  
where $V$ is the volume of the fluid motion. Unlike the previous integrals it is an inviscid invariant \cite{lt}-\cite{l} as the helicity itself (the conservation of the helicty can be related, by the Noether's theorem, to the fundamental relabeling symmetry \cite{mor}-\cite{pm}). It follows from the Eq. (8) that the Levich-Tsinober integral can have a non-zero value even when the average helicity is equal to zero. \\

  Since the helicity itself is an inviscid invariant, as well as the kinetic energy, the Kolmogorov's scaling phenomenology for the inertial range of scales \cite{my} was applied to helicity in the Ref. \cite{fr} and kinetic energy spectrum $E(k) \propto \varepsilon_h^{2/3} k^{-7/3}$ (where $\varepsilon_h =|d\langle h \rangle/dt|$) was obtained instead of the Kolmogorov spectrum $E(k) \propto \varepsilon^{2/3} k^{-5/3}$ (where $\varepsilon =|d\langle {\bf u}^2 \rangle/dt|$). At this approach the average helicity and kinetic energy are considered as adiabatic invariants in the inertial range of scales. This approach can be also applied to the Levich-Tsinober integral as an inviscid invariant. However, one should take into account that unlike energy and helicity, which are quadratic invariants, the Levich-Tsinober integral is a quartic invariant (see Eq. (8)). Therefore, one should use $\varepsilon_I =|dI^{1/2}dt|$ in order to apply the Kolmogorov's scaling phenomenology to this case:
$$
E(k) \propto \varepsilon_I^{2/3} k^{-4/3}  \eqno{(9)}
$$

\section{Distributed chaos}

  Statistically stationary homogeneous and isotropic turbulence for incompressible fluids is usually numerically simulated in a cubic volume by the Navier-Stokes equations:
$$
 \frac{\partial {\bf u}}{\partial t} = - {\bf u} \cdot \nabla {\bf u} 
    -\frac{1}{\rho} \nabla {\cal P} + \nu \nabla^2  {\bf u} + {\bf f} \eqno{(10)}
$$
$$ 
\nabla \cdot {\bf u} = 0\eqno{(11)}
$$
with periodic boundary conditions (where ${\bf u}$ is a velocity field, ${\cal P}$ is a pressure field, $\nu$ is a viscosity, ${\bf f}$ is a forcing). Well defined broadband kinetic energy spectra are observed in these direct numerical simulations (DNS) already for small values of the Taylor-Reynolds number $Re_{\lambda} = 8$ (see, for instance, Ref. \cite{pky}). Since one cannot expect a turbulent motion at such values of $Re_{\lambda}$ \cite{sreeni},\cite{sb} these spectra can be attributed to a chaotic motion of the fluid.\\ 

  Figure 1 shows, in the log-log scales, a kinetic energy spectrum obtained in a direct numerical simulation reported in the Ref. \cite{pky} (the spectral data were taken from Fig. 2 of the Ref. \cite{pky}). The dashed curve is drawn in the figure to indicate exponential spectral decay
$$
E(k) = a \exp-(k/k_c)   \eqno{(12)}
$$ 
The dotted arrow indicates position of the wavenumber $k_c$ (the wavenumbers are normalized by the Kolmogorov scale \cite{my} $\eta = (\nu^3/\varepsilon)^{1/4}$). The exponential spectral decay is a well known (but not well understood) feature of the chaotic motions of fluids and plasmas (see, for instance, Refs. \cite{fm}-\cite{b1} and references therein).  \\

     Increase of  $Re_{\lambda}$ results in more complex chaotic motion which is characterized by an ensemble with the statistically varying $k_c$ and $a$ parameters, and the ensemble average should be used in order to compute the spectral decay
$$
E(k) = \int P(a,k_c) ~\exp-(k/k_c)~ dadk_c  \eqno{(13)}
$$      
with $P(a,k_c)$ as a joint probability distribution for the ensemble parameters $a$ and $k_c$. For statistically independent parameters $a$ and $k_c$
$$
E(k) \propto \int P(k_c) ~\exp-(k/k_c)~ dk_c  \eqno{(14)}
$$
\begin{figure} \vspace{-1.1cm}\centering
\epsfig{width=.45\textwidth,file=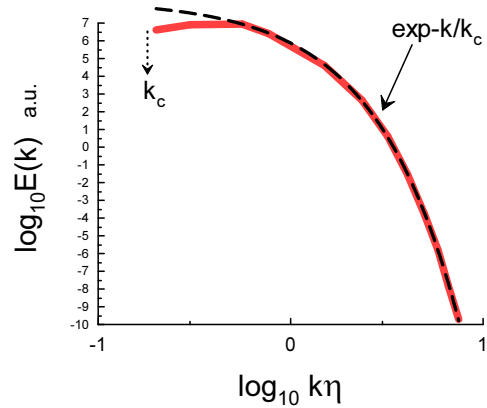} \vspace{-4.5cm}
\caption{Kinetic energy spectrum for isotropic homogeneous motion of incompressible fluid at $Re_{\lambda} = 8$.} 
\end{figure}

  One can consider a stretched exponential spectrum  
$$
E(k) \propto \int P(k_c) ~\exp-(k/k_c)~ dk_c \propto \exp-(k/k_{\beta})^{\beta}  \eqno{(15)}
$$     
as a natural generalization of exponential spectrum.\\

  Asymptotic behaviour of the probability density $P(k_c)$ at large values of $k_c$ can be immediately inferred from Eq. (15) \cite{jon}
$$
P(k_c) \propto k_c^{-1 + \beta/[2(1-\beta)]}~\exp(-bk_c^{\beta/(1-\beta)})  \eqno{(16)}
$$
  
 Then assuming a scaling of the characteristic velocity $v_c$ at large values of the $k_c$
$$  
v_c \propto k_c^{\alpha}  \eqno{(17)}
$$
and normal (Gaussian) distribution of the $v_c$ we obtain from the Eq. (16) relationship between $\alpha$ and $\beta$
$$
\beta = \frac{2\alpha}{1+2\alpha}   \eqno{(18)}
$$
\begin{figure} \vspace{-1.9cm}\centering
\epsfig{width=.42\textwidth,file=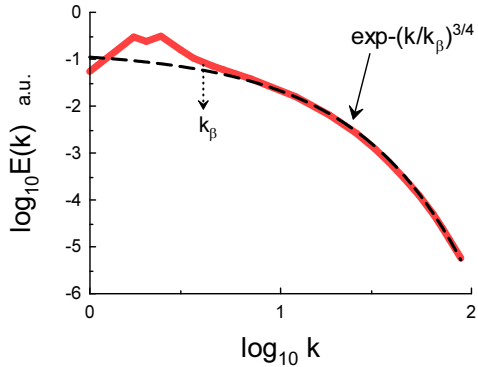} \vspace{-3.6cm}
\caption{Kinetic energy spectrum for isotropic homogeneous {\it non-helical} turbulence at $Re_{\lambda} = 100$.} 
\end{figure}

\begin{figure} \vspace{-0.5cm}\centering
\epsfig{width=.42\textwidth,file=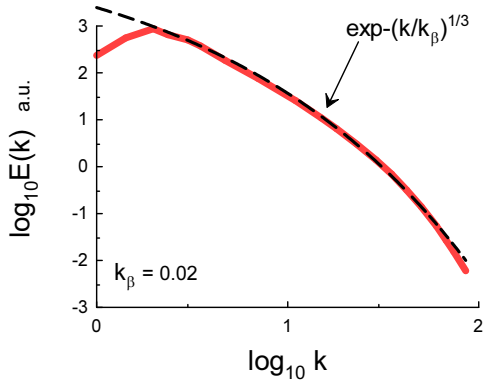} \vspace{-3.77cm}
\caption{As in the Fig. 2 but for the turbulence with strong multiscale helical injection.} 
\end{figure}

\begin{figure} \vspace{-0.5cm}\centering
\epsfig{width=.42\textwidth,file=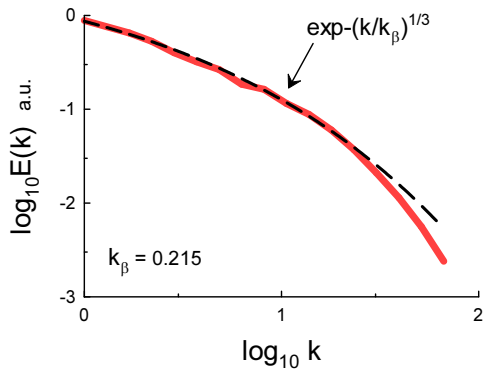} \vspace{-3.8cm}
\caption{Kinetic energy spectrum of the helical isotropic turbulence with multiscale Langevin stirring applied at all wavenumbers up to the dissipative scales.  } 
\end{figure}
  
  Using the dimensional considerations we obtain from the Eqs. (2) and (17)
$$
v_c \propto \lvert\mathcal{S}\rvert^{1/2} k_c^{3/2}     \eqno{(19)}
$$
i.e. $\alpha =3/2$ for the Birkhoff-Saffman turbulence. Then it follows from the Eqs. (15) and (18) that for this turbulence 
$$
E(k) \propto \exp-(k/k_{\beta})^{3/4}   \eqno{(20)}
$$
(cf. the Ref. \cite{b1}).

   Analogously for turbulence dominated by the Chkhetiani invariant Eq. (5)
$$
v_c \propto \lvert\mathcal{C}\rvert^{1/2} k_c     \eqno{(21)}
$$
and, correspondingly
$$
E(k) \propto \exp-(k/k_{\beta})^{2/3},   \eqno{(22)}
$$
whereas for the turbulence dominated by the Levich-Tsinober invariant Eq. (8)
$$
v_c \propto \lvert I \rvert^{1/4} k_c^{1/4}     \eqno{(23)}
$$
and, correspondingly
$$
E(k) \propto \exp-(k/k_{\beta})^{1/3}.   \eqno{(24)}
$$

\section{Direct numerical simulations}

  In recent DNS reported in Ref. \cite{kes} the Navier-Stokes equations Eqs. (10-11) were numerically solved using two types of forcing: one - with energy injection applied at a large-scale ($k_f \simeq 2.2$) and another - with helicity injection applied at all scales (multiscale) belonging to the supposed inertial range. Both forcing funtions ${\bf f}({\bf x},t)$ were divergence free and delta-correlated in time. 
  
    The kinetic energy spectrum, obtained in the DNS for the first type of forcing (non-helical turbulence) at a statistically steady state, is shown in figure 2 (the spectral data were taken from Fig. 1 of the Ref. \cite{kes}, $Re_{\lambda} =100$). The dashed curve is drawn in the Fig. 2 to indicate the stretched exponential spectral decay Eq. (20) corresponding to the Birkhoff-Saffman (non-helical) turbulence. The dotted arrow indicates position of the wavenumber $k_{\beta}$.
\begin{figure} \vspace{-1.7cm}\centering
\epsfig{width=.42\textwidth,file=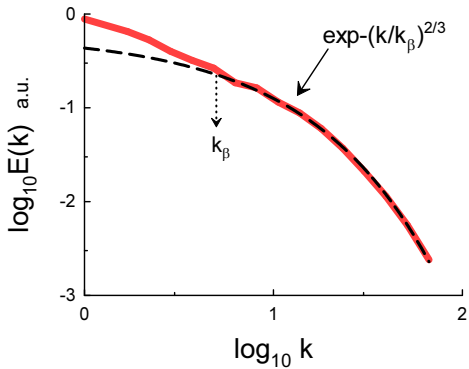} \vspace{-3.9cm}
\caption{The same spectral data as in the Fig. 4 but with the Chkhetiani invariant approximation Eq. (22) for the near-dissipation range. } 
\end{figure}
\begin{figure} \vspace{-0.5cm}\centering
\epsfig{width=.42\textwidth,file=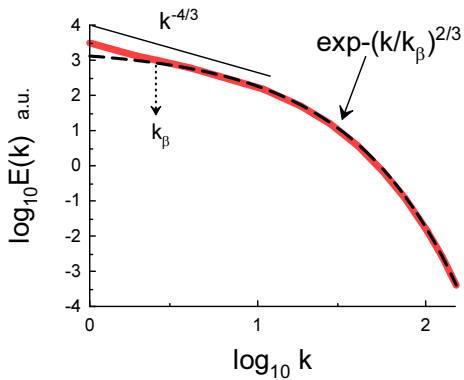} \vspace{-3.7cm}
\caption{Kinetic energy spectrum (spherically integrated) for the helical turbulence forced by the Euler large-scale forcing scheme. } 
\end{figure}
\begin{figure} \vspace{-0.5cm}\centering
\epsfig{width=.42\textwidth,file=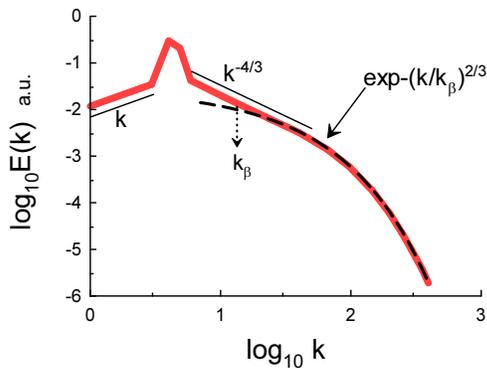} \vspace{-4.3cm}
\caption{Kinetic energy spectrum for the narrow-band forced helical turbulence. } 
\end{figure}
    Figure 3 shows the kinetic energy spectrum, obtained in the DNS for the second type of forcing (maximally helical turbulence) at a statistically steady state (the spectral data were also taken from Fig. 1 of the Ref. \cite{kes}). The dashed curve is drawn in the Fig. 3 to indicate the stretched exponential spectral decay Eq. (24) corresponding to the Levich-Tsinober invariant dominated helical turbulence. \\
    
    In another recent DNS \cite{bif} a strong multiscale Langevin stirring was used for the helical injection in an isotropic homogeneous turbulence. The helical forcing was a Gaussian white-in-time with a power-law spectrum and energy injection was applied at all wavenumbers up to the dissipative scales. Figure 4 shows the kinetic energy spectrum, obtained in the DNS (the spectral data were taken from Fig. 2b of the Ref. \cite{bif}). The dashed curve is drawn in the Fig. 4 to indicate the stretched exponential spectral decay Eq. (24) corresponding to the Levich-Tsinober invariant dominated inertial range of the helical turbulence. 
    
    Figure 5 shows the same spectral data but the dashed curve is drawn in the Fig. 5 to indicate the stretched exponential spectral decay Eq. (22) corresponding to the Chkhetiani invariant dominated near-dissipation range of the helical turbulence (let us recall that the Chkhetiani invariant, as well as the Loitsianskii and the Birkhoff-Saffman integrals are invariants of the {\it viscous} Navier-Stokes equations \cite{otto1},\cite{otto2}). The dotted arrow indicates position of the wavenumber $k_{\beta}$. Naturally, the Chkhetiani invariant dominated range is overlapped with that dominated by the Levich-Tsinober invariant in this case.\\

    The interplay of the Levich-Tsinober and the Chkhetiani turbulence can be also seen in the results of a recent DNS reported in the Ref. \cite{val}. In this DNS an Euler large-scale forcing was applied in order to obtain a helical homogeneous turbulence. This type of forcing resembles truncated Euler dynamics \cite{cic}: the lowest wavenumber (forced) modes, belonging to a sphere $0 \leq |{\bf k}| \leq k_f$, are obeying the incompressible ideal Euler equation, while being independent of the other modes. The modes with $|{\bf k}| > k_f$ obey the incompressible Navier-Stokes equations, while being dependent on the modes inside the Euler (forcing) sphere. The inviscid invariants of the ideal Euler equation are conserved in this truncated system. Solenoidal random velocity fields with a given kinetic energy spectrum were used as initial conditions ($Re_{\lambda} = 136$ and $k_f = 1.5$). 

    Figure 6 shows the kinetic energy spectrum (spherically integrated), obtained in the DNS. The spectral data were taken from Fig. 3b (the case $c_{h}^1$) of the Ref. \cite{val}. The dashed curve is drawn in the Fig. 6 to indicate the stretched exponential spectral decay Eq. (22) corresponding to the Chkhetiani invariant dominated near-dissipation range of the helical turbulence (the dotted arrow indicates position of the wavenumber $k_{\beta}$), whereas the straight line is drawn for reference of the scaling law Eq. (9) corresponding to the Levich-Tsinober invariant dominated inertial range. And again (as in the previous case) the two ranges are overlapped.  \\
    
    In recent DNS Ref. \cite{al} the large-scale helical forcing was applied at wavenumbers $k_f \leq |{\bf k}| \leq k_{f}+1$ with $k_f =4$. While the forcing amplitude was fixed its Fourier modes phases were randomly varied. Figure 7 shows the kinetic energy spectrum obtained in the DNS (the spectral data were taken from Fig. 2b of the Ref. \cite{al}). The dashed curve is drawn in the Fig. 7 to indicate the stretched exponential spectral decay Eq. (22) corresponding to the Chkhetiani invariant dominated near-dissipation range of the helical turbulence (the dotted arrow indicates position of the wavenumber $k_{\beta}$), whereas the straight lines are drawn for reference of the scaling law Eq. (9) corresponding to the Levich-Tsinober invariant dominated inertial range and to the low-wavenumber scaling Eq. (6) corresponding to the Chkhetiani invariant.

\section{Laboratory experiments behind multiscale grids}

    It is commonly believed that turbulent flows behind mechanical grids can provide laboratory simulation of isotropic (in a wide sense - including the reflection symmetry) homogeneous turbulence (see, for instance, Ref. \cite{tm} and references therein). However, direct measurements of helicity in the turbulence behind the grids indicate violation of the reflection symmetry \cite{kit}. Moreover, the Levitch-Tsinober integral Eq. (8) can have a finite (non-zero) value even when the average helicity is eqial to zero. Therefore, helicity related effects can play a significant role in the turbulent flows with zero average helicity, especially for the multiscale forcing (see above). In recent years the multiscale grids were actively used in the laboratory experiments and it is interesting to look at results obtained in these experiments. \\
    
    Figure 8 shows the one-dimensional (longitudinal) kinetic energy spectrum obtained in a laboratory experiment \cite{kro} with a multiscale grid at distance $x/M_1=20$ behind the grid (where $M_1$ is the largest  mesh size of the grid). The Reynolds number based on the mesh size $M_1$ is $Re = 6 \times 10^4$. The figure corresponds to the Fig. 8 of the Ref. \cite{kro}. The forcing geometric scales are uniformly distributed over the range of energy containing scales. The dashed curve is drawn in the Fig. 8 to indicate the stretched exponential spectral decay Eq. (24) corresponding to the Levich-Tsinober invariant dominated turbulence. One can see that although there were no special attempts to inject helicity in the flow the helicity related Levich-Tsinober invariant apparently plays a significant role in the inertial range of scales. It can be considered as a kind of reflection symmetry breaking, even if the mean helicity is still negligible in this turbulent flow (cf. Refs. \cite{lt},\cite{l},\cite{bt} and references therein). \\
    
    Figure 9 shows the one-dimensional (longitudinal) kinetic energy spectrum obtained in a laboratory experiment \cite{hl} with a multiscale ('fractal') grid at distances $x/M=25,~35,~44.6$ behind the grid (where $M$ is the largest  mesh size of the grid, $R_{\lambda} \simeq$ 94, 86, 81 correspondingly). The normalization used in the Fig. 9 results in a single (collapsed) spectrum for the three distances from the grid (the spectral data were taken from the Fig. 9b of the Ref. \cite{hl} and $\lambda$ is the Taylor microscale \cite{my}). The dashed curve is drawn in the Fig. 9 to indicate the stretched exponential spectral decay Eq. (24) corresponding to the Levich-Tsinober invariant dominated turbulence (the straight line corresponds to the scaling (9) in the large-scale part of the inertial range).

\begin{figure} \vspace{-1.5cm}\centering
\epsfig{width=.42\textwidth,file=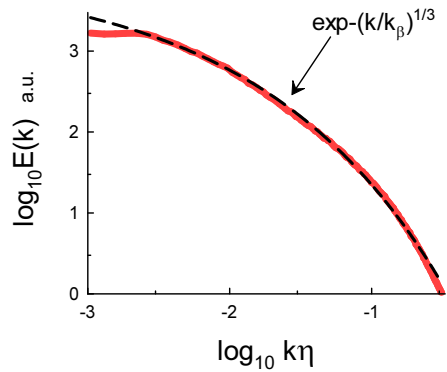} \vspace{-3.9cm}
\caption{One-dimensional (longitudinal) kinetic energy spectrum obtained in a laboratory experiment behind a multiscale grid. } 
\end{figure}
\begin{figure} \vspace{-0.3cm}\centering
\epsfig{width=.42\textwidth,file=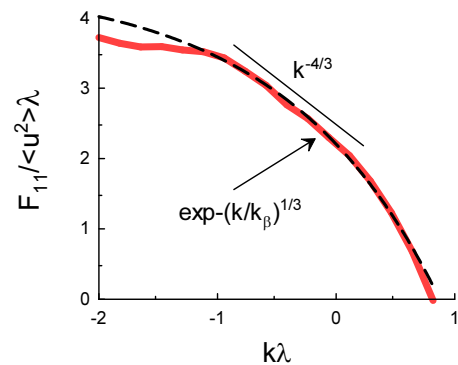} \vspace{-4cm}
\caption{Normalized one-dimensional (longitudinal) kinetic energy spectrum obtained in a laboratory experiment behind a multiscale ('fractal') grid.} 
\end{figure}

\section{Acknowledgement}

I thank O.G. Chkhetiani for sending his papers and comments, E. Levich for discussion and P.-A. Krogstad for sharing his data.

\end{document}